\documentclass[12pt,preprint]{aastex}
\usepackage{amsmath}	% Advanced maths commands
\usepackage{amssymb}	% Extra maths symbols
\newcommand{\kms}{\,km\,s$^{-1}$}
\newcommand{\gcm}{\,g\,cm$^{-1}$}
\newcommand{\ergs}{\,erg\,s$^{-1}$}
\newcommand{\ha}{H$\alpha$}
\newcommand{\he}{He\,I 5876\,\AA}
\newcommand{\fex}{[Fe\,X] 6374\,\AA}

\begin{document}

\title{Wind of presupernova IIn SN~1997eg}
\bigskip

\author{N.N.~Chugai
}
\affil{Institute of astronomy RAS, Moscow 119017, Pyatnitskaya 48\\
\textup{nchugai@inasan.ru} 
}

%======================================================

\begin{abstract}

Spectra and phototometry of type IIn supernova SN 1997eg are used to determine 
properties of the circumstellar gas lost by the presupernova during the latest 200 years 
before the explosion. The analysis of narrow \ha\ and \fex\ results in the wind velocity 
 $u = 20$\kms, significantly lower than the earlier accepted value (160\kms) 
upon the bases of the 
radial velocity of a blue absorption wing of the narrow \ha. That high velocity of the 
wind in our picture is related to the preshock gas accelerated by the cosmic ray precursor.
The modelling of the circumstellar interaction results in the wind density parameter 
$\dot{M}/u$ that being combined with the wind velocity 
suggests the presupernova  mass loss rate of 
$1.6\cdot10^{-3}\,M_{\odot}$\,yr$^{-1}$.
The wind density is consistent with the \fex\ luminosity. The model \ha\ luminosity  
also agrees with the observational value. Recovered wind properties indicate that 
the presupernova at the final evolutionary stage was a massive red supergiant with a high mass loss rate, but not the LBV-supergiant as suggested earlier.

%\keywords{stars: massive --- supernovae: general --- supernovae:
%individual (\objectname{SN 1997eg}) --- cosmic rays}
\end{abstract}

\section{Introduction}

Supernova SN 1997eg (Nakano \& Aoki 1997) belongs to type IIn (Filippenko \& Barth 1997) 
with signatures of a dense circumstellar matter (CSM). The luminosity of 
most SNe~IIn both in lines and continua is due to the dissipation of a kinetic energy 
in forward and reverse shocks in the course of the SN/CSM interaction. Likewise in the canonical SN~1988Z (Filippenko 1991, Statakis \& Sadler 1991), strong emission lines of SN~1997eg  
show three components (Hoffman et al. 2008): the narrow lines of the CSM, the broad component 
(FWHM $\approx 10^4$\kms) and the intermediate component (FWHM $\approx 2000$\kms). 
The broad component originates from the undisturbed SN envelope and the cool dense shell 
(CDS) between forward and reverse shocks. The intermediate component originates from dense 
CS clouds interacting with the forward shock (Chugai \& Danziger 1994).

The origin of SNe IIn, particularly, the main-sequence mass and mass loss 
mechanism, are far from clear. In this regard of great importance are 
properties of the 
CSM, i.e., the velocity, density and spatial scale of CSM distribution, which could 
shed light on the issue.
It is noteworthy, the stellar wind is not the only potential source of the dense CSM.
In some cases the CSM can be a result of a mass ejection in a 
single explosive perturbation likewise in SN 2006gy (Woosley et al. 2007).
Insight into the CSM density can be obtained via the modelling of the bolometric luminosity and 
the evolution of the supernova expansion velocity. The later is recovered from 
emission line profiles. 
As to the bolometric luminosity of SN 1997eg the situation is hampered by the 
absence of long term observations. 
Yet several photometric observations around day 90  
 (Tsvetkov \& Pavlyuk 2004) in combination with estimates of the supernova velocity 
 could be sufficient for the determination of the CSM density.

With the known density and wind velocity one immediately recovers the 
mass loss rate. In most SN IIn however the low spectral resolution does not permit to infer the  
wind velocity. In this regard of particular significance is the spectrum of 
SN 1997eg on 198 day with the resolution of 7\kms (Salamanka et al. 2002).
Yet the former interpretation of narrow lines is doubtful. 
The point is that apart from narrow emission lines with FWHM of 30-40\kms\ the 
spectrum shows broad absorption component of \ha\ with the radial velocity 
-160\kms\ in the blue wing, which is attributed to the wind velocity. 
The ambiguity of the issue of the wind velocity is a strong motivation to revisit 
the interpretation of SN~1997eg observations and reassess CSM parameters.
It should be emphasised that the wind velocity is crucial for the problem of the 
SN~IIn origin. The velocity of 10-30\kms\ would indicate that a presupernova 
before the explosion was a red supergiant (RSG), whereas the wind velocity of
$\sim200$\kms\ would imply that 
presupernova could be a luminous blue variable (LBV). The latter are  actually 
associated with some SNe~IIn, e.g., SN~2009ip (Pastorello et al. 2013).

The primary goal of the paper is the determination of the presupernova
wind velocity and density. The presentation starts with the description and realization 
of the model for the emission line profiles of \ha\ and \he. As a result we recover the velocity 
of the undisturbed supernova envelope at the reverse shock; this velocity will be 
used then for the determination of the wind density in the SN/CSM interaction model. 
To recover the wind velocity I propose the model for the narrow \ha\ formed in the 
the slow wind with the preshock acceleration by the cosmic ray (CR) precursor. 
In addition, the wind velocity and gas temperature will be recovered in the region 
of the coronal \fex\ line emission. The SN/CSM interaction then will be modelled to 
derive the wind density. In the last section I discuss results and make  
suggestions on the nature of the presupernova.

Following Hoffman et al. (2002) the SN 1997eg distance of 40 Mpc is adopted.
The explosion moment is rather uncertain. However high expansion velocity and 
their fast decrease indicate that supernova was detected soon after the explosion. 
Therefore in accord with Hoffman et al. (2002) we adopt that the supernova 
exploded 9 days before the discovery.

\section{Broad emission lines and supernova expansion velocity}
\label{sec:vbroad}

The supernova expansion in a dense CSM, created by the presupernova wind, 
brings about forward and reverse shocks. Between them a 
cool ($T \sim 10^4 K$) dense shell forms, since the reverse shock is strongly radiative;
outer shock can be partially radiative or adiabatic depending on the wind density.
The supernova deceleration manifested in the decrease 
of the width of broad line wings can be used to recover the wind density.
The appropriate decceleration indicator is the maximum velocity 
of the undisturbed supernova envelope ($v_{sn}$) at the reverse shock that 
can be estimated from the maximum velocity in the blue wing of
the H$\alpha$ emission line, since the red wing is affected by the 
Thomson scattering.
Yet the velocity jump between supernova and dense shell ($v_{sn} > v_{ds}$) 
can result in the blue wing enhancement due to Thomson scattering and 
the photon blueshift into the radial velocities $< -v_{sn}$ 
(an analogue of the Fermi acceleration). 
Therefore, a reliable method for the supernova velocity determination should be 
based on the modelling of the profiles of broad emission lines. 

A general model for the formation of emission lines in SNe~IIn is lacking 
though such a model is unlikely plausible given highly heterogeneous SNe~IIn family. 
Yet a source of the emission is undoubtedly the dissipation 
of the kinetic energy in shocks. As already mentioned, strong emission lines, 
particularly, H$\alpha$ in the SN~1997eg spectrum are composed by three 
components: broad, narrow, and intermediate (Hoffman et al. 2008). The 
narrow component originated in the CSM is irrelevant to the issue of supernova 
expansion velocity, so it will be addressed in a separate section. However, 
it should be noted that at the early stage the narrow line emitted by the CSM 
may acquire broad wings formed due to the multiple Thomson scattering of 
the line photons in the optically thick circumstellar cocoon as occured in 
early SN 1998S (Chugai 2001). For SN 1997eg, however this effect is 
negligible at the considered late stages.

The schematic picture of the supernova interaction with the CSM showing  
zones for the emission of the broad, narrow, and intermediate components is 
presented in Fig. 1. It should be mentioned that in fact the broad component 
consists of two componens: the emission of the undisturbed supernova envelope 
and of the dense shell residing between two shocks. The modelling of the 
line profile from the undisturbed supernova ejecta with the kinematics 
of $v = r/t$ and adopted distribution of the emissivity versus radius is a rather 
simple problem. More complicated model however is needed to describe the broad line 
originated from the dense shell, the issue addressed below.

\subsection{Emission line from dense shell}
\label{sec:cds}

In one-dimensional approach the CDS is very thin in accord with 
the momentum flux conservation 
$\Delta R/R \sim (v_{th}/v_{ds})^2\sim 10^{-6}-10^{-5}$ 
(where $v_{th} \sim 10$\kms\ is the thermal velocity of the CDS). There are two 
problems with the line emission from that thin shell. The radiation of the 
optically thick line, e.g. H$\alpha$, escapes the thin CDS in the surface regime. 
The line width in the comoving frame for the optical depth $<10^5$ is 
$\Delta \nu \lesssim 6v_{th}/\lambda$. Even for  
the thermalized line source function the expected H$\alpha$ luminosity from the CDS 
$L_{32} = 4\pi^2 R_{ds}^2B_{\nu}(T)\Delta \nu$ is relatively low. 
Another problem is that the optically thick line in 
that case has an M-shaped profile (Gerasimovi\v{c} 1933), which is not the case  
for SN 1997eg.

In reality, the CDS is liable to the strong deformation due to the 
Rayleigh-Taylor instability (Chevalier \& Blondin 1995, 
Blondin \& Ellison 2001), which results in the penetration of spikes of 
dense cool gas into the hot gas of the forward shock wave 
followed by mixing and formation of a  
complicated three-dimensional structure of the cool dense gas in the form 
of corrugated two-dimensional elements. This structure  
with a low volume filling factor resides in a spherical layer 
of the width of $\Delta R/R \sim 0.1$ (Blondin \& Ellison 2001).
The contact surface area $S$ of the cool dense gas in this case 
significantly exceeds the initial surface area $4\pi R^2$, so the area 
ratio $A = S/(4\pi R^2) \gg 1$.
The formation of the optically thick line emitted by the cool dense gas in this case 
was considered earlier and the main result is that for the area ratio $A \gg 1$ 
the line luminosity gets larger, while the line profile is significantly modified compared to the case of spherical thin shell (Chugai et al. 2004).
Hereafter when talking about the line from the CDS we mean the line 
from the strongly deformed CDS.

In the analytic model (Chugai et al. 2004) the distribution of the CDS material is described by an ensemble of random thin disks. 
Each disk is considered as a particle that can emit and absorb resonance radiation.
Transport of the radiation in this shell is treated in the Sobolev approximation.
The probability for the photon escape in the direction defined by 
cosine $\mu$ of the angle between wave vector and the radius is 
\begin{equation}
\beta = (1 - \mbox{e}^{-Q})/Q, \qquad Q = Q_0/(1 - \mu^2) \,.
\label{eq:beta}
\end{equation}
A zero velocity radial gradient is adopted, which is a good approximation. 
The parameter $Q_0$ depends on area ratio $A$, thermal width of the 
local line absorption coefficient, relative width of the mixing layer
$\Delta R/R$ and the line optical depth of a single disk. 
The propability $\beta$ actually is a macroscopic analogue of the Sobolev escape probability.
The effect of the $Q_0$ value on the optically thick line is as follows: 
for $0.1 < Q_0 < 0.5$ the profile is flat-topped, while for 
$Q_0 \gtrsim 1$ the profile becomes parabolic.

\subsection{Intermediate component}
\label{sec:inter}

The intermediate component, that contributes significantly to H$\alpha$,
originates from the 
interaction of the forward shock with dense CS clouds (Chugai \& Danziger 1994).
The characteristic cloud shock speed $v_c$ is determined by the density contrast of 
the cloud relative to the intercloud gas $\chi = \rho_c/\rho_{ic}$ and the forward 
postshock velocity that is comparable to the velocity of the CDS $v_{ds}$:
\begin{equation}
v_c \approx 2v_{ds}/\sqrt{\chi}\,,
\end{equation}
where the prefactor of 2 takes account of the four times larger postshock density.
For $\chi \sim 10^2$ and $v_{ds} = 5000$\kms\ the expected cloud 
shock speed is $v_c\sim 10^3$\kms, a typical value for the intermediate component.
Higher velocities in the wings of the intermediate component are the outcome of 
the acceleration of cloud fragments in the post-forward shock flow. The maximal velocities of cloud fragments turn out to be comparable to the postshock 
velocity of forward shock (Klein et al. 1994). One should bear in mind that 
the cloud shock is essentially radiative, so cloud fragments are rather 
dense and cool ($T \sim 10^4$\,K).

The modelling of the intermediate component requires setting the
 emissivity distribution versus velocity and 
radius, $j(v,r)$. I consider the homogeneous source distribution 
along the radius and the parametric form for $j(v)$ on the interval  
$v_{min} < v < v_{max}$, with $v_{min} \sim v_c$ and $v_{max} \lesssim v_{ds}$:
\begin{equation}
j(v) \propto
 \left\{\begin{array}{ccc}
  \dfrac{v-v_{min}}{v_c-v_{min}}\,,  & \quad \textrm{if $v_{min}<v<v_c$}\,, \\
  \left[\dfrac{v_{max}-v}{v_{max}-v_c}\right]^q\,,   & \quad \textrm{if $v_c<v<v_{max}$}\,, \\
    0\,, & \quad \textrm{otherwise\,.}
    \end{array}\right.
\label{eq:vpdf}
\end{equation}
The power index $q$ in this expression is close to unity. The function $j(v)$ has a maximum 
at the typical velocity of the cloud shock $v = v_c$ that is a free parameter.
The range $v_{c,min} < v < v_c$ where $j(v)$ grows is introduced to take into account the scatter 
of clouds properties. Without this detail the profile would be flat-topped
which is not observed.
We adopt $v_{min} = 0.9v_c$. The radial distribution is set to be uniform in the range 
$\delta R \sim 0.2R$. Somewhat similar form of $j(v,r)$ was used earlier 
(Chugai 2009) for the line profiles in SN 2006jc.

\subsection{H$\alpha$, He\,I 5876 \AA\ and velocity of supernova envelope}
\label{sec:mbroad}

The model for the broad lines includes three line-emitting zones:  (1) homologously expanding supernova envelope ($v = r/t$, $r < R_0 = 1$), 
(2) the disturbed CDS ($R_0 < r < R_1 = 1.1$) with the average velocity $v_{ds}$ and random velocities in the range $\pm 0.05v_{ds}$ (Blondin \& Elison 2004),
 and (3) the cool dense  gas of fragmented CS clouds in the forward shock ($R_0 < r < R_2 = 1.2$). 
First two zones are responsible for the broad component, while the third one for the 
intermediate component. We neglect the thickness of the reverse shock and adopt the radius 
of the undisturbed supernova envelope to be equal $R_0$. The relative contribution 
of the supernova envelope into the line emission is $f_{sn}$, the contribution of the 
CDS is $f_{ds}$, and the rest fraction $1 - f_{ds} - f_{sn}$ is contributed by the intermediate component.

The Monte Carlo technique is employed for the radiation transfer computation.
The model profiles of H$\alpha$ and He\,I 5876 \AA\ are compared to the observed 
spectra for three epochs, 57, 80, and 204 days after the discovery in Figures 2, 3, and 4.
Observed spectra are taken from the database {\it Weizmann supernova data repository} 
(Yaron \& Gal-Yam 2012). Model parameters are presented in Table, which contains starting from the first column: 
the moment after the discovery, velocity of the undisturbed supernova envelope, CDS velocity, 
maximal velocity of the gas responsible for the intermediate component, 
power index in the 
expression (\ref{eq:vpdf}), parameter $Q_0$ in equation (\ref{eq:beta}), 
optical depth of 
the undisturbed supernova and the CDS in continua, single scattering albedo in ejecta and the CDS, contribution of ejecta and CDS to H$\alpha$ and He\,I 5876 \AA. The 
adopted CS cloud shock speed $v_c$ is 850\kms\ on days 57 and 80, and 920\kms\ on day 204.

On days 57 and 80 the broad component is the same for the He\,I 5876 \AA\ and
H$\alpha$ lines; in both cases the homogeneous source distribution in ejecta is adopted. 
On day 204 for He\,I 5876 \AA\ the line emissivity in supernova is concentrated 
towards rhe center $j \propto 1/v$, 
whereas for H$\alpha$ as before $j = const$. Remarkably,  
the intermediate component of He\,I 5876 \AA\ is significantly weaker 
than in H$\alpha$. This probably reflects a low excitation degree  
in the dense CS cloud fragments responsible for the intermediate component. 
On day 204 the components are not easily distiguished visually, so in this case 
profiles of all the components are shown. 

The primary goal of the line profile modelling is the velocity 
of the undisturbed supernova at the reverse shock (Table, second column).
The relative uncertainty of this value is 10\%. It is instructive to compare 
these values to the visual measurements of the maximal velocity in the 
H$\alpha$ blue wing. On days 57, 80, and 204 thus estimated values are equal  
to 10000\kms, 8500\kms, and 7000\kms, respectively, in agreement with 
values in Table. The uncertainty of both the CDS velocity  and $v_{max}$ is 15\%. Other parameters are determined with 
the uncertainty of 20\% and these parameters fortunately do not affect inferred 
supernova velocities.

\section{Narrow lines and wind velocity}

The high resolution spectrum of SN 1997eg (7\kms) on day 198 (Salamanka et al. 2002) 
shows along with narrow emission lines (FWHM 30-40\kms) also  the H$\alpha$ 
absorption component 
with the radial velocity in the blue wing of -160\kms. It is natural to account for that high velocity in the same way as suggested for 
the broad absorption component of narrow H$\alpha$ of SN~1998S (type IIL/n) 
(Chugai et al. 2002).
The central point of the interpretation of the absorption profile 
is that the high velocity is attributed to the preshock gas accelerated by 
the CR precursor that is a feature of the diffusive shock acceleration 
mechanism (Druri \& F\"{o}lk 1981).

The CR pressure at the shock ($p_c$) required to provide the preshock gas acceleration up to $u = 160$\kms\ can be estimated from the equation of the momentum flux conservation. 
The integration in the preshock region $-\infty < x < 0$ gives: 
  \begin{equation}
  (p + \rho v^2)_0 - (p + \rho v^2)_{-\infty} =0\,.
  \label{eq:precur}
  \end{equation}
Neglecting the gas density variation in the preshock and using sensible  approximations $v_{\infty} \approx v_{ds}$, 
$p_{-\infty} = 0$, $p_0 = p_c$, $v_0 = v_{ds} - u$, and making use of $u/v_{ds} \ll 1$ one gets from equation (\ref{eq:precur}) $p_c/\rho v_{fs}^2 = 2(u/v_{fs})$. 
For $u = 160$\kms\  and $v_{fs} = 5000$\kms\ we infer the ratio of the CR 
pressure to the upstream gas dynamic pressure to be $p_c/\rho v_{fs}^2 \approx 0.06$. 
  This is rather modest requirement given claims that up to 50\% of the postshock pressure in supernova remnants may be due to the CR (Helder et al. 2009).
 
Preempting the H$\alpha$ modelling it is instructive to derive velocity of the undisturbed wind using  
the strongest coronal line of [Fe\,X] 6374\,\AA\ in the same spectrum.
In Figure 5 the observed [Fe\,X] 6374\,\AA\ line (Salamanka et al. 2002) 
is shown with the overplotted model. For stationary wind the best fit is found for the 
wind speed of 19.7\kms\ and the Doppler broadening for $T = 0.9\cdot10^6$ K. The accuracy of 
the derived values is 10\%. The inferred temperature is close to the value of 
$1.1\cdot10^6$ K that corresponds to the maximum of the relative abundance 
of Fe\,X ion (Burgess \& Seaton 1964) in a coronal approximation.

The kinematic model we use for the H$\alpha$ suggests that the wind velocity decreses from 
160\kms\ to 20\kms\ in the radius range of $1.2 < r < 1.4$ and remains constant futher on. The region of the H$\alpha$ formation is bounded by the radius $1.6R_0$, because anyway the population 
of the second level decreases with the radius due to the Ly$\alpha$ escape. 
In the optimal model the second level population smoothly drops by a factor of 1.7 
in the range $1.2 < r < 1.6$. We adopt 
a thermal Doppler broadening for the local absorption coefficient 
$k = k_0\exp(-[\Delta \lambda/\Delta \lambda_D]^2)$, where  $\Delta \lambda$ is the 
wavelenth displacement from the central wavelength, and $\Delta \lambda_D = \lambda_0(v_{th}/c)$ is the Doppler width, $v_{th} = (2k_BTN_A)^{1/2}$, $k_B$ is the Boltzmann constant,
$T$ is the gas temperature, $N_A$ is the Avogadro number. The scattering in the 
line frequences is assumed to be conservative with the complete frequency redistribution. As a background radiation 
for narrow line we consider emission in the broad and intermediate components of 
the H$\alpha$.
In the ejecta and CDS we adopt the same extinction as in the model on day 204.
The optimal model (Fig. 5) reproduces the observed narrow H$\alpha$ 
for the wind optical depth in H$\alpha$ of 0.4, wind velocity of 20\kms, and thermal velocity 
of hydrogen of 30\kms.  The latter corresponds to the gas temperature of 
$5.4\cdot10^4$\,K.

The major result of this modelling is that the H$\alpha$ 
narrow emission component forms primarily in the wind with the velocity of 20\kms, 
which supports the result of the [Fe\,X] 6374\,\AA\ modelling.
In this regard a question arises on the difference between 
the temperature derived from  H$\alpha$ ($5\cdot10^5$\,K) and from 
[Fe\,X] 6374 \AA\ ($9\cdot10^5$\,K). 
A possible explanation is that in the inner wind, where the H$\alpha$ forms, the 
temperature is indeed lower than in the outer wind, where the coronal lines form.
The reason could be a thermal instability, when a sharp transition between 
 cold and hot states is plausible for similar specific 
 heating rates similar to the sharp transition between chromosphere and corona.

 \section{Supernova deceleration and wind density}

The interaction of the supernova with the CSM is modelled in the thin shell approximation 
(Chevalier 1982, Guiliani 1982). I recap only essential points, since 
the model has been described earlier (e.g. Chugai et al. 2004). 
The gas swept up by the forward and reverse shock forms the shell that 
is considered to be thin ($\delta r/r \ll 1$). Its motion is governed by the equation of motion and equation 
of the mass conservation. The optical bolometric luminosity is adopted to be equal to 
the luminosity of both shocks. The shock luminosity at the moment $t$ is calculated as the shock 
kinetic luminosity multiplied by the radiation efficiency $\eta = t/(t+t_c)$, where 
$t_c$ is the cooling time of the postshock gas for the density four times of the preshock 
density. We omit the radiation diffusion effect and the luminosity powered by 
the internal energy of the supernova explosion. 
The density distribution in the supernova ejecta is set as 
$\rho = \rho_0/[1 + (v/v_0)^8]$, where $\rho_0$ and $v_0$ are specified by the ejecta mass $M$ 
and kinetic energy $E$. The adopted density distribution is very close to that of 
hydrodynamic models of a SNe~IIP.

The high expansion velocity of the SN 1997eg envelope on day 57 ($\approx10^4$\kms) 
indicates a very weak decceleration of outer layers. This in turn suggests that 
in the inner region the density is significantly lower than that extrapolated inward according to the steady wind ($\rho \propto r^{-2}$). Indeed, our simulations show 
the law $\rho \propto r^{-2}$ in the region $r < 4\cdot10^{13}$ cm 
does not permit to describe both the luminosity and the velocity 
evolution. The successful model (Fig. 6)
suggests the CSM with the homogeneous density 
for $r < r_b = 4\cdot10^{15}$ and $\rho \propto r^{-2}$ for $r > r_b$. 
The shown estimate of the bolometric luminosity is obtained using photometry by 
Tsvetkov \& Pavlyuk (2004). 
The adopted supernova ejecta mass in this model is $15~M_{\odot}$. With this 
choice the optimal kinetic energy is $2.2\cdot10^{51}$ erg. The recovered wind density 
parameter in the range $r \geq r_b$ is $w = \dot{M}/u = 5\cdot10^{16}$\gcm. 
With the found wind density and the wind velocity of 20\kms\ the presupernova 
mass loss rate turns out to be $1.6\cdot10^{-3}\,M_{\odot}$\,yr$^{-1}$.
It should be 
emphasised that the found wind density is insensitive to the adopted ejecta mass. 
E.g., the model with $M = 1~M_{\odot}$ and $E = 4.5\cdot10^{50}$ erg reproduces 
observation as good as the $15~M_{\odot}$ model for the same wind density.
The reason for the model degeneracy with respect to the mass is related to the power law distribution of supernova density $\rho(v)$ in outer layers that can be 
the same for the different appropriate values of $M$ and $E$.

\section{Discussion}

The wind density parameter and the wind velocity (20\kms) are the major results of 
the new interpretation of the available observational data on SN 1997eg.
These parameters imply the presupernova mass loss rate of 
$1.6\cdot10^{-3}\,M_{\odot}$\,yr$^{-1}$. The CDS radius on day 213   
is $1.26\cdot10^{16}$ cm. The wind material 
with the velocity of 20\kms\ at this radius was lost 200 years before the explosion.
Remarkably, the new estimate of the wind velocity is significantly (8 times) 
lower than the previusly adopted value of 160\kms. 

The central point of our interpretation of narrow lines in the spectrum of 
SN1997eg is the conjecture that high velocities detected in the absorption blue wing 
of narrow H$\alpha$ are due to the wind acceleration in the CR precursor of the shock wave. This suggestion is in line with the 
earlier interpretation of a similar phenomenon in the H$\alpha$ absorption of 
SN~1998S (Chugai et al. 2002). Thus at present we have two cases of SN IIn that 
demonstrate in the H$\alpha$ a signature of the preshock wind acceleration by the
CR precursor. There is a suspicion that in the high resolution spectrum 
of SN 2002ic (SN IIn) on day 265 (Kotak et al. 2004) we see a similar phenomenon: 
the H$\alpha$ narrow emission has a width FWHM of 80\kms, whereas the 
absorption blue wing extends to -250\kms. The preshock acceleration in the 
CR precursor is observed in the Tycho supernova remnant as well. Lee et al. (2007) 
detected gas velocities in the preshock region in the range of 60-130\kms\ that 
have been attributed to the gas acceleration in the CR precursor. This range of 
velocities corresponds to the ratio 0.03 - 0.06 with respect to the shock velocity. 
With this ratio extended to the SN 1997eg one expects to find preshock 
velocities of 150 - 300\kms\ in line with the observed value of 160\kms.

Let us turn now to implications of the inferred CR pressure for the 
the radio emission from SN 1997eg. 
The radio flux from SN 1997eg at 3.6 cm on days 177 and 186 was detected at the 
level of 0.5 mJy (Lassey \& Weiler 1998). This flux corresponds to the 
radio luminosity $\nu L_{\nu} = 8.4\cdot10^{36}$ erg s$^{-1}$. The 
intercloud wind with the coronal temperature of $10^6$\,K and the shock 
radius of $R = 1.24\cdot10^{16}$ cm suggests the opticall depth to the 
$ff$-absorption of 1.45, so the unabsorbed radio luminosity should be of 
$3\cdot10^{37}$ erg s$^{-1}$. 

We assume that the width of the radio-emitting shell is $\Delta R = 0.1R$ and 
it is filled with relativistic electrons with the energy density 
$\epsilon_e = 3K_{ep}p_c = 1.26\cdot10^{-2}$ erg cm$^{-3}$, where we adopt 
the standard electron-to-proton ratio $K_{ep} = 10^{-2}$, the found ratio 
$p_c/\rho v_{fs}^2 = 0.06$,  with $v_{fs} = 5200$\kms\ and inferred wind density.
The Lorentz factor of radio-emitting electrons at 3.6 cm is $\gamma = 55B^{-0.5}$. 
For the standard energy spectrum of relativistic electrons $dn/dE \propto E^{-2}$ in 
the range $E_1 < E < E_2$ the radio luminosity is 
$\nu L_{\nu}  = \epsilon_e V/(2t_s\ln{(E_2/E_1)})$,where $V$ is the volume of the radio-emitting shell, $t_s = 3m_ec^2/(4\sigma_TcW_m\gamma)$ is the characteristic 
time of synchrotron losses, and $W_m = B^2/8\pi$. For sensible energy range 
of accelerated electrons $E_1 = 1$ MeV, $E_2 = 10^3$ MeV one gets 
$\nu L_{\nu} = 1.3\cdot10^{38}B^{3/2}$ erg s$^{-1}$. 
The magnetic field required to account for the unabsorbed 
radio luminosity is thus $B \approx 0.4$ G. Remarkably, this value is in the 
range of  0.1 - 0.6 G inferred for 12 radio supernovae through synchrotron 
self-absorption effects (Chevalier (1998). The seemingly large 
value of the magnetic field of SN 1997eg thus turns out quite reasonable.

It is instructive to compare the model and observational H$\alpha$ luminosities 
in order to verify the consistency of the broad line modelling. On day 204 the model 
suggests that the Thomson optical depth of the unshocked ejecta is 
$\tau_T = 0.6$. For the CDS radius at this age of $r = 1.26\cdot10^{16}$ cm 
one gets the average electron number density $n_e = 7.2\cdot10^7$ cm$^{-3}$. 
Assuming the solar composition and the H$\alpha$ effective recombination coefficient 
$\alpha_{32} = 3.4\cdot10^{-13}$ cm$^3$ c$^{-1}$ for $T_e = 5000$\,K 
(Osterbrock 1989) one obtains the H$\alpha$ recombination luminosity of the unshocked ejecta 
$L_{sn} = 4.4\cdot10^{40}$\,erg s$^{-1}$. Given the contribution of unshocked 
ejecta to the overall H$\alpha$ of 0.27 we come to the total H$\alpha$ luminosity 
in our model of $1.6\cdot10^{41}$\,erg\,s$^{-1}$ that coinsides with the observed 
H$\alpha$ luminosity of $1.7\cdot10^{41}$\,erg\,s$^{-1}$ (Salamanca et al. 2002).

To which extent the inferred wind density is 
consistent with other observational data? One can find the model luminosity 
of the coronal [Fe\,X] 6374 \AA\ line on day 198 and compare it to the observed 
luminosity of $1.5\cdot10^{38}$\ergs\ (Salamanka et al. 2002) for the distance of 
40~ Mpc.
The maximum fraction of Fe\,X ion to iron ($f_{10} = 0.33$) is attained for the 
temperature of $\approx 10^6$\,K (Burgess \& Seaton 1964). The collisional excitation 
coefficient for that temperature is $3.8\cdot10^{-9}$ cm$^3$\,s$^{-1}$ 
(Aggarval \& Keenan 2005). The emission measure in the region 
$r > 1.45\cdot10^{16}$ cm for the wind density parameter 
$w = 1.6\cdot10^{-3}\,M_{\odot}$\,yr$^{-1}$ assuming the solar composition 
reproduces the observed [Fe\,X] 6374 \AA\ luminosity, 
if the mass fraction of the hot intercloud wind is $f_{ic} = 0.6$. The 
latter is realistic value and the agreement between the model and 
observations thus turns out acceptable.

Another estimate of the wind density can be obtained from the ratio of 
nebular to auroral lines of O\,III.
Taking into account errors of the determination of observed fluxes of [O\,III] lines 
in the noisy spectrom on day 198 (Salamanca et al. 2002, Figure 3) the ratio 
$F(5007+4959)/F(4363)$ turns out to be in the range 0.5-1. For the temperature of $10^5$\,K 
corresponding to the maximum fraction of the  O\,III ion (B\"{o}hringer 1998) the 
found ratio suggests the electron number density of 
$(0.75-1.7)\cdot10^7$\,cm$^{-3}$ (Osterbrock 1989).
This range is in accord with the preshock 
electron number density $n_e = 1.4\cdot10^7$\,cm$^{-3}$ implied by the inferred 
wind density.

The recovered wind velocity ($20$\kms) implies that the presupernova of SN 1997eg 
was RSG, not LBV supergiant as earlier suggested on the basis of the  
wind velocity of 160\kms\ (Salamanka et al. 2002, Hoffman et al. 2008). 
Some SNe~IIn indeed are associated with LBV presupernovae , e.g., SN~2009ip, 
in which case the wind velocity is of 240\kms\ (Pastorello et al. 2013), 
and the luminous presupernova is identified with the massive LBV  
$M_{ms} \gtrsim 60~M_{\odot}$  (Foley et al. 2011). 
The insight into the initial mass of SN~1997eg 
is provided by the sample of 255 RSG in M31 (Massey \& Evans 2016). This 
sample lies on the HR-diagram below the evolutionary track for $37~M_{\odot}$, 
which implies the SN 1997eg progenitor mass $< 40~M_{\odot}$. 
The mass lower limit should be around $25~M_{\odot}$ since according 
to a general wizdom stars from the mass range of $9 - 25~M_{\odot}$ explode as 
SNe~IIP (Woosley et al. 2002). It is remarkable that the relation 
between the mass loss rate and RSG luminosity, viz.,  
$\dot{M} = 10^{-4}(L/10^5L_{\odot})^{2.1}$ (Salasnich et al. 1999) combined 
with the mass loss rate of 
$1.6\cdot10^{-3}\,M_{\odot}$\,yr$^{-1}$ suggests the SN~1997eg 
progenitor of $30\,M_{\odot}$, in accord with the estimated 
range of $25-40\,M_{\odot}$.

 \section{Conclusion}
 
 The interpretation and modelling of the available spectral and photometric data on 
 type IIn supernova SN 1997eg permits us to infer major parameters of the 
 presupernova mass loss during the last 200 yr before the explosion: 
 the wind velocity of 20\kms  and mass loss rate of 
 $1.6\cdot10^{-3}\,M_{\odot}$\,yr$^{-1}$.
 The recovered wind density is consistent with the value independently estimated 
 from the flux of the coronal line [Fe\,X] 6374\,\AA.
 The wind properties indicate that the presupernova at the final evolutionary 
 stage was a red supergiant with the large mass loss rate. The progenitor 
 mass at the main sequence propably was in the range of $25-40\,M_{\odot}$.

\clearpage   
%****************************************************************
%
\begin{table*}
\centering 
\caption[]{Parameters of models for the H$\alpha$ and He\,I 5876\,\AA\ lines}
\vspace{0.5 cm}
\label{tab:param}
\begin{tabular}{l|c|c|c|c|c|c|c|c|c|c|c|c|c}
\hline
 Days & $v_{sn}$ & $v_{ds}$  & $v_{max}$ & $q$  & $Q_0$ & $\tau_{sn}$ 
  & $\tau_{ds}$ & $\omega_{sn}$ & $\omega_{ds}$ & $f_{sn}$ & $f_{ds}$ & $f_{sn,He}$ & $f_{ds,He}$ \\
\hline
 57    &  9500   &   7000 &   5000   &  1.5    &  0.5  &  0.9  &
         0.13    &   0.4  & 0.4   &  0.5    &  0.1  & 0.66 & 0.3 \\
    
 80    &  8500   &   6100  &   4500   &  1.1    &  0.7   &  0.7  &
         0.1     &   0.4  & 0.4   &  0.42  &  0.19  & 0.7  & 0.25 \\  
 
 204  &  7100   &   5200  &  4300    &  1    &  1      &  0.6  &
        0.05    &   1    & 0.3  &  0.27   &  0.33   & 0.68 & 0.3 \\   
\hline
\end{tabular}
\label{tab:mod}
\end{table*}
%

%==================================================================================

\clearpage
%xxxxxxxxxxxxxxxxxxxxxxxxxxxxxxxxxxxxxxxxxxxxxxxxxxxxxxxxxxxxxx
\begin{figure}[h]
\includegraphics[trim=100 100 0 0,  width=0.9\textwidth]{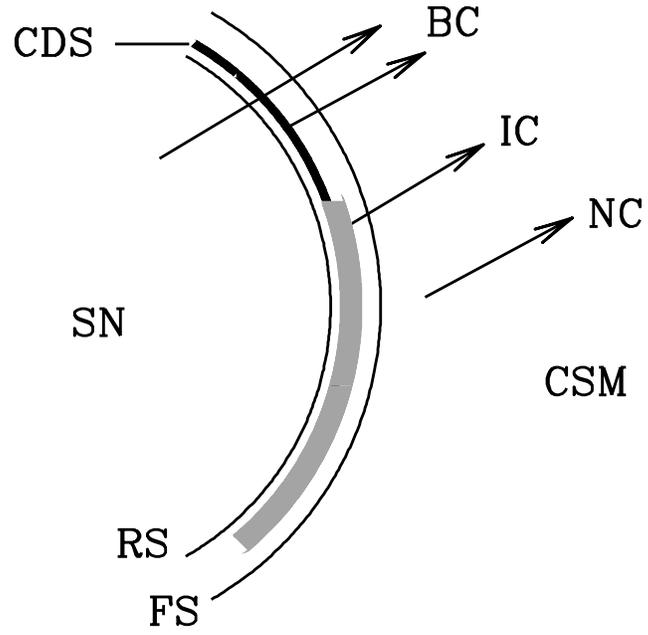}
	\caption{
	 Schematic picture of the supernova (SN) interaction with the 
circumstellar matter (CSM), that shows regions of the formation of 
three components of an emission line: narrow component (NC), broad component (BC), 
and intermediate component (IC). Two thin arc lines correspond to the forward 
shock (FS) and reverse shock (RS); thick line shows the cool dense shell (CDS);
grey arc shows a zone which contains cool dense gas of shocked and 
fragmented circumstellar clouds; this layer is responsible for the intermediate 
component.
	}
	\label{fig:cart}
\end{figure}
%========================================================
%xxxxxxxxxxxxxxxxxxxxxxxxxxxxxxxxxxxxxxxxxxxxxxxxxxxxxxxxxxxxxx
\clearpage
%xxxxxxxxxxxxxxxxxxxxxxxxxxxxxxxxxxxxxxxxxxxxxxxxxxxxxxxxxxxxxx
\begin{figure}[h]
\includegraphics[trim=0 0 0 0,  width=0.9\textwidth]{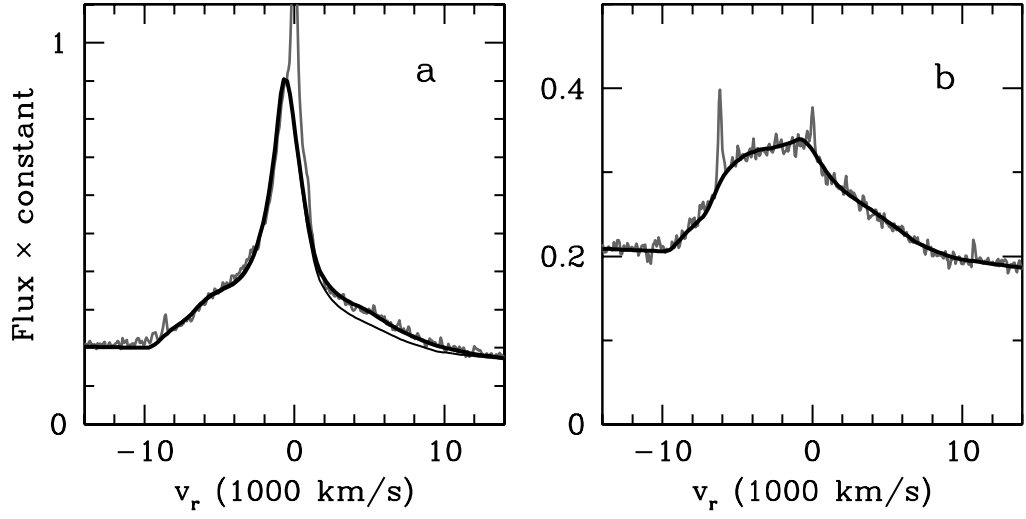}
	\caption{
	H$\alpha$ (panel {\bf a}) and He\,I 5876\,\AA\ (panel {\bf b}) lines in the 
SN~1997eg spectrum on day 57 compared to models 
(cf. Table). Thin line on the panel {\bf a} shows the model without the 
contribution of He\,I 6674\,\AA.	
	}
	\label{fig:sp1}
\end{figure}
%========================================================
%xxxxxxxxxxxxxxxxxxxxxxxxxxxxxxxxxxxxxxxxxxxxxxxxxxxxxxxxxxxxxx
\clearpage
%xxxxxxxxxxxxxxxxxxxxxxxxxxxxxxxxxxxxxxxxxxxxxxxxxxxxxxxxxxxxxx
\begin{figure}[h]
\includegraphics[trim= 0 0 0 0,  width=0.9\textwidth]{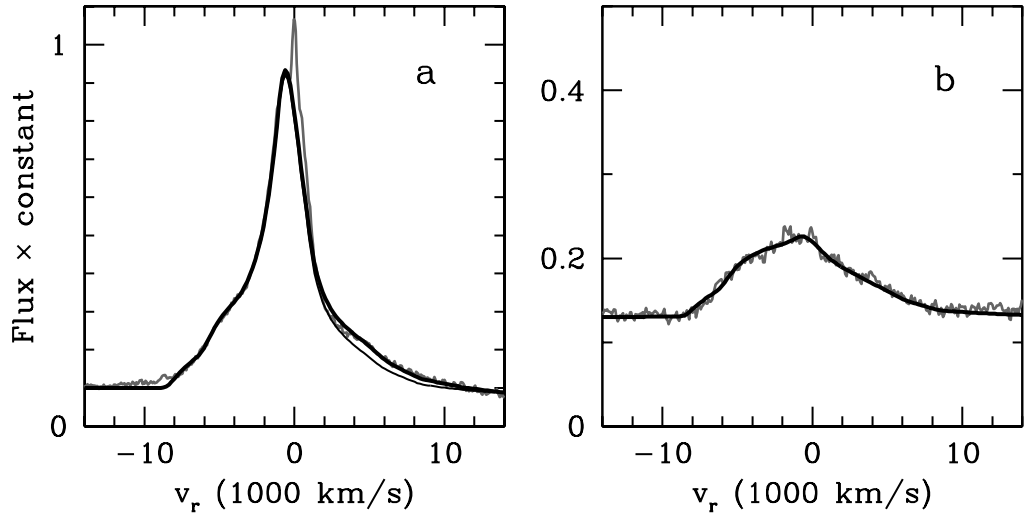}
	\caption{
	The same as Fig. 2 but for day 80.
	}
	\label{fig:sp2}
\end{figure}
%========================================================
%xxxxxxxxxxxxxxxxxxxxxxxxxxxxxxxxxxxxxxxxxxxxxxxxxxxxxxxxxxxxxx
\clearpage
%xxxxxxxxxxxxxxxxxxxxxxxxxxxxxxxxxxxxxxxxxxxxxxxxxxxxxxxxxxxxxx
\begin{figure}[h]
\includegraphics[trim=0 0 0 0,  width=0.9\textwidth]{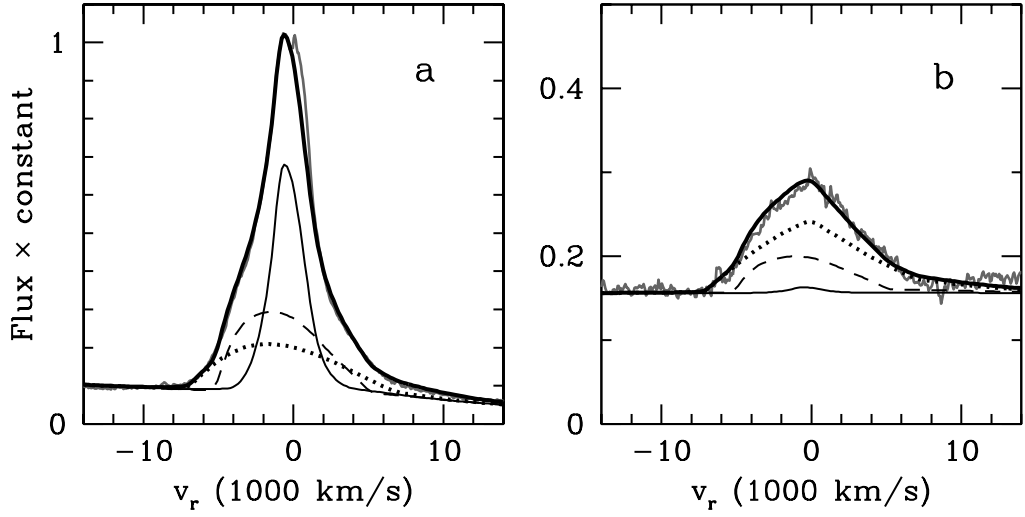}
	\caption{ 
	The same as Fig. 2 but for day 204. Additionally shown are 
  the intermediate component ({\em thin solid line}), the contribution 
   of the dense shell ({\em dashed line}), and the contribution of 
   undisturbed supernova ejecta ({\em dotted}).	
	}
	\label{fig:sp3}
\end{figure}
%========================================================
%xxxxxxxxxxxxxxxxxxxxxxxxxxxxxxxxxxxxxxxxxxxxxxxxxxxxxxxxxxxxxx
\clearpage
%xxxxxxxxxxxxxxxxxxxxxxxxxxxxxxxxxxxxxxxxxxxxxxxxxxxxxxxxxxxxxx
\begin{figure}[h]
\includegraphics[trim=0 0 0 0,  width=0.9\textwidth]{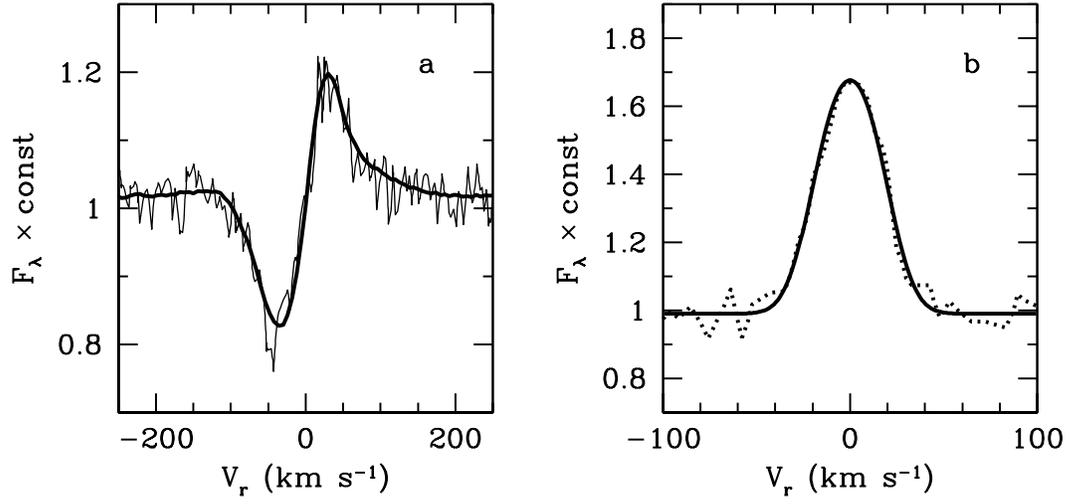}
	\caption{
	Circumstellar H$\alpha$ ({\bf a}) and [Fe\,X] 6374\,\AA\ ({\em dotted}) 
in the SN 1997eg spectrum on day 198 compared to models ({\em thick lines}).	
	}
	\label{fig:nar}
\end{figure}
%========================================================
%xxxxxxxxxxxxxxxxxxxxxxxxxxxxxxxxxxxxxxxxxxxxxxxxxxxxxxxxxxxxxx
\clearpage
%xxxxxxxxxxxxxxxxxxxxxxxxxxxxxxxxxxxxxxxxxxxxxxxxxxxxxxxxxxxxxx
\begin{figure}[h]
\includegraphics[trim=0 0 0 0,  width=0.9\textwidth]{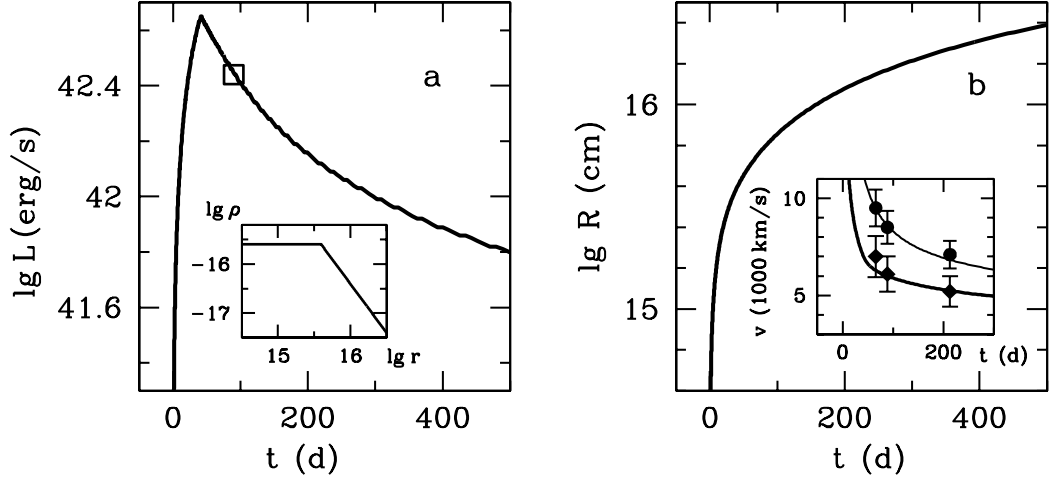}
	\caption{
	Bolometric light curve (panel {\bf a}) and radius of the dense 
shell (panel {\bf b}) in the model of $15~M_{\odot}$ supernova ejecta interaction with the CSM. In the panel {\bf a} the square is the bolometric luminosity according 
to photometry data (Tsvetkov \& Pavlyuk 2004). The inset in panel {\bf a} 
shows the density distribution of the CSM in the model. 
The inset in panel {\bf b} shows the model supernova velocity at the reverse shock
({\em thin line}) compared to the velocity estimates from the H$\alpha$ 
({\em filled circles}) and the model velocity of the dense shell compared to 
estimates from the H$\alpha$ ({\em diamonds}).	
	}
	\label{fig:dyn1}
\end{figure}

\end{document}